\documentclass[prl,aps,twocolumn]{revtex4}
\usepackage{amsmath}
\usepackage{amsfonts}
\usepackage{amssymb}
\usepackage{graphicx}
\usepackage{color}

\newcommand{\be}{\begin{equation}}
\newcommand{\ee}{\end{equation}}
\newcommand{\bea}{\begin{eqnarray}}
\newcommand{\eea}{\end{eqnarray}}

\input{tcilatex}
\begin{document}

\title{Quantum spin metal state on a decorated honeycomb lattice}
\author{K. S. Tikhonov$^{1,2}$ and M. V. Feigel'man$^{1,2}$}
\affiliation{$^1$ L. D. Landau Institute for Theoretical Physics, Kosygin str.2, Moscow
119334, Russia}
\affiliation{$^2$ Moscow Institute of Physics and Technology, Moscow 141700, Russia}
\date{\today }

\begin{abstract}
We present a modification of exactly solvable spin-$\frac{1}{2}$ Kitaev
model on the decorated honeycomb lattice, with a
ground state of "spin metal" type. The model is diagonalized in terms of
Majorana fermions; the latter form a 2D gapless state with a Fermi-circle
those size depends on the ratio of exchange couplings. Low-temperature heat
capacity $C(T)$ and dynamic spin susceptibility $\chi (\omega ,T)$ are
calculated in the case of small Fermi-circle. Whereas $C(T)\sim T$ at low
temperatures as it is expected for a Fermi-liquid, spin excitations are gapful
and $\chi (\omega ,T)$ demonstrate unusual behaviour with a power-law peak near the resonance
frequency. The corresponding exponent as well as the peak shape are
calculated.
\end{abstract}

\maketitle

Quantum spin liquids ~\cite{PWA1,PWA2,Wen,rev1,rev2} present very
interesting examples of strongly correlated phases of matter which do not
follow the classical Landau route: no local order parameter is formed while
the entropy vanishes at zero temperature. Although quite a number of
different proposals for the realization of spin liquid states are available,
and many interesting results were obtained numerically, see e.g.~\cite{rev1}%
, the progress in analitical theory was hindered for a long time due to the
absense of an appropriate quantum spin model exactly solvable in more than
one spatial dimension. Seminal results due to A. Kitaev~\cite{Kitaev06} may
pave the way to fill this gap. Kitaev proposed spin-$\frac{1}{2}$ model with
anisotropic nearest-neighbour spin interactions $J_{\alpha }\sigma
_{i}^{\alpha }\sigma _{j}^{\alpha }$ (where $\alpha =x,y,z$) on the
honeycomb lattice. The model allows for the two-dimensional generalization
of the Jordan-Wigner spin-to-fermion transformation and can be thus exactly
diagonalized. Kitaev model realizes different phases as its ground-states,
the most interesting of them corresponds to the region around symmetric
point $J_{x}=J_{y}=J_{z}=J^{\prime }$. The excitations above this
ground-state constitute a single branch of massless Dirac fermions (Kitaev
presents them in terms of Majorana fermions those spectrum contains two
symmetric conical points). Thus the Kitaev honeycomb model presents exactly
solvable case of the \textit{critical} QSL. Although long-range spin
correlations vanish in this model presisely due to its integrability~\cite%
{Baskaran07}, it presents valuable starting point for the construction of
analitically controllable theories possessing long-range spin correlations.

\begin{figure}[tbp]
\includegraphics[width=8cm,height=5cm]{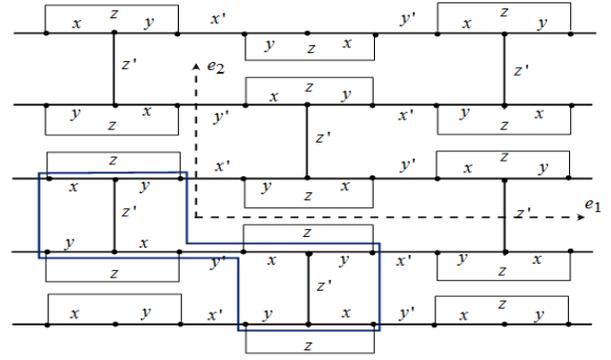}
\caption{(Color online) "Brick-wall" representation of the decorated honeycomb lattice.
Elementary cell for the Majorana Hamiltonian contains 12 sites and is bounded by the blue line.}
\label{pcell}
\end{figure}

Still, here we discuss another extension of the model~\cite{Kitaev06}: it is
interesting to find exact realizations of other types of QSLs, the one with 
\textit{gapful} excitation spectrum, and the one with a whole surface of
gapless excitations~\cite{spinmetal}. Gapful QSL was recently found by Yao
and Kivelson~\cite{YK}. They proposed specific generalization of the Kitaev
model, where each site of the honeycomb lattice is replaced by a triangle
(we will call it, for brevity, "3-12 lattice"), with internal coupling
strengths equal to $J$ and inter-triangle couplings equal to $J^{\prime}$.
Topologically equivalent structure of such a lattice is shown in Fig.~\ref{pcell}.
Yao-Kivelson model is  exactly solvable and
contains a critical point at $g\equiv J^{\prime }/J=\sqrt{3}$. At $g=\sqrt{3}
$ the excitation spectrum has single low-energy branch of Dirac fermions,
like the original Kitaev model, whereas at any other $g$ the excitation
spectrum is gapful. Yao and Kivelson have shown that the ground state at $g<%
\sqrt{3}$ is a topologically nontrivial chiral spin liquid with Chern number 
$\mathrm{C}=\pm 1$, whereas $g>\sqrt{3}$ phase is topologically trivial,
with $\mathrm{C}=0$. Exactly solvable QSL model of spin-metal type  
with spins-$\frac{3}{2}$ was proposed by Yao, Zhang and Kivelson~\cite{YZK}.
In the present Letter we take a different route: we show that slight
modification of the Hamiltonian of spin-$\frac{1}{2}$ Yao-Kivelson model on
the 3-12 lattice leads to the spin-metal QSL with a pseudo-Fermi-circle.
Similar approach was proposed recently in Ref.~\cite{Baskaran09} where QSL
with pseudo-Fermi-line was found in the spin-$\frac{1}{2}$ model on a
decorated square lattice. Apart from another lattice studied, our study
differs from~\cite{Baskaran09} in two respects: i) our Fermi-liquid-like
ground state is the result of spontaneous symmetry breaking leading to a
"chiral antiferromagnet" ordering defined in terms of 3-spin products, ii)
we present analitic results for heat capacity and dynamic spin
susceptibility in the limit of small Fermi-surface size. 
We consider the Hamiltonian: 
\begin{equation}
\mathcal{H}=\sum_{l=\left\langle ij\right\rangle }J_{l}\left( \mathbf{\sigma 
}_{i}\mathbf{n}_{l}\right) \left( \mathbf{\sigma }_{j}\mathbf{n}_{l}\right)
+\lambda \sum_{x^{\prime },y^{\prime },z^{\prime }-links}T_{p}T_{p^{\prime
}}.  \label{H}
\end{equation}%
Here $T_{p}=\sigma _{ia}^{x}\sigma _{jb}^{y}\sigma _{kc}^{z}$ is the
three-spin "exchange" operator corresponding to the $p$-th triangle. We take 
$J_{x,y,z}=J,$ $J_{x^{\prime },y^{\prime },z^{\prime }}=J^{\prime }$ and
without loss of generality assume $~J,~J^{\prime }>0$. Unit vectors $\mathbf{n}_l$
are parallel to $x$, $y$ and $z$ axis for the corresponding links $x$, $x^\prime$, etc.
Eq. (\ref{H}) reduces to the original Hamiltonian of Ref.~\cite{YK} at $\lambda =0$. \
This spin Hamiltonian is rather special, since it posesses large number of
independent integrals of motion, so called fluxes defined as $%
W_{p}=\prod_{s=1}^{n}\sigma _{j_{s}}^{\alpha }\sigma _{j_{s-1}}^{\alpha }$,
where $j_{0},~j_{1},~...,~j_{n}=j_{0}$ defines a minimal close loop on the
lattice. All $W_{p}$ commute with Hamiltonian and with each other and divide
total Hilbert space into sectors, corresponding to different sets of $W_{p}$
eigenvalues. 
\begin{figure}[tbp]
\includegraphics[width=8cm,height=2.5cm]{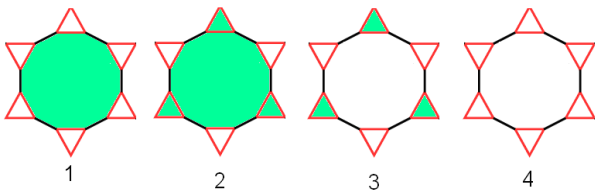}
\caption{(Color online) Flux configurations in $G_{1,2,3,4}$ states
correspondingly. Empty/filled plaquette corresponds to the flux $+1,-1$ for
dodecagons and $+i,-i$ for triangles.}
\label{pfluxes}
\end{figure}
Hamiltonian (\ref{H}) can be diagonalized by means of Kitaev representation
of spins via 4 Majorana fermions: $\sigma _{i}^{\alpha }=ic_{i}c_{i}^{\alpha
}$, where four Majorana operators $c_{i},c_{i}^{x},c_{i}^{y},c_{i}^{z}$ are
defined on each site of the lattice and satisfy anticommutation relations $%
\left\{ c_{i}^{\alpha },c_{j}^{\beta }\right\} =2\delta _{ij}\delta _{\alpha
\beta }$. This way, each spin (2-dim Hilbert space) is represented by four
Majoranas\, (4-dim Hilbert space). This is  a representation in the
extended Hilbert space; all physical states   should
satisfy the constraint: $D_{i}\left\vert \Psi _{phys}\right\rangle =\left\vert \Psi
_{phys}\right\rangle $ for any lattice site $i$, where $%
D_{i}=c_{i}c_{i}^{x}c_{i}^{y}c_{i}^{z}$. Operators $D_{i}$ is the gauge transformation
operator for the group $\mathcal{Z}_{2}$. Hamiltonian (\ref{H}),
extended to the Hilbert space of Majorana fermions, reads ($%
u_{ij}=ic_{i}^{\alpha }c_{j}^{\alpha }$): 
\begin{equation}
\mathcal{H}=-i\sum_{l=\left\langle ij\right\rangle
}J_{l}c_{i}u_{ij}c_j+\lambda \sum_{x^{\prime },y^{\prime },z^{\prime
}-links}t_{p}t_{p^{\prime }}  \label{H2}
\end{equation}%
with $t_{p}=u_{ab}u_{bc}u_{ca}$. Note, that $[u_{ij},H]=0$ and $u_{ij}^{2}=1$
. These integrals of motion are gauge-dependent; they are related to the
gauge-invariant fluxes $W_{p}=\prod_{s=1}^{n}\left(
-iu_{j_{s}j_{s-1}}\right) $. There are two types of $W_{p}$\, : 1) fluxes,
corresponding to the triangular loops $W_{p}^{(3)}=\pm i$, and 2) fluxes $%
W_{p}^{(12)}=\pm 1$ corresponding to the dodecagon loops. They respond
differently to the time reversal transformation: ${\mathcal{\hat{T}}}%
W_{p}^{(12)}=W_{p}^{(12)}$, but ${\mathcal{\hat{T}}}W_{p}^{(3)}=-W_{p}^{(3)}$%
. It was shown in Ref.~\cite{Yao} that the ground state of Hamiltonian (\ref%
{H}) with $\lambda =0$ corresponds to all $W_{p}^{(12)}=-1$ and all $%
W_{p}^{(3)}~$ are equal (either to $i$, or to $-i$)\,; these two global eigenstates are
related by the $\mathcal{\hat{T}}$ \ inversion. We show now that in some range
of couplings $J,J^{\prime }$ even very small $\lambda $ stabilizes another
type of the ground state, with variables $W_{p}^{(3)}=\pm i$ ordered
alternatively (like in the AFM Ising model on honeycomb lattice), and with
Fermi-line of gapless excitations.

Hamiltonian (\ref{H2}) can be diagonalized for any periodic configuration of
the gauge field $u_{ij}$. However, we restrict our consideration to the
states with the same flux periodicity as the original lattice. Thus we are 
left with 4 gauge-nonequivalent states $G_{1,2,3,4}~$ shown in Fig.~\ref%
{pfluxes} , plus their time-reversal
partners $G_{i^{\prime }}={\mathcal{\hat{T}}}G_{i}$. Since the gauge
field $u_{ij}$ which correspond to the flux configurations $G_{2,4}$ does not
fit into 6-site unit cell, in order to describe all states $G_{1..4}$ we use
elementary cell containing $12$ sites, shown in Fig~\ref{pcell}, with $%
\mathbf{e}_{1}=\mathbf{e}_{x}$ and $\mathbf{e}_{2}=\mathbf{e}_{y}$
(hereafter lengths are  measured in units of the lattice spacing).

After the gauge is fixed, we are left with the  Hamiltonian $\mathcal{H}$,
restricted to the Majorana space, and denoted by $H$ . 
This Hamiltonian can be diagonalized in
terms of Fourier-transformed Majorana fields $\psi _{\alpha ,\mathbf{k}}=%
\frac{1}{\sqrt{2N}}\sum_{\mathbf{r}}e^{-i\mathbf{kr}}c_{\alpha ,\mathbf{r}}$%
, where subscript $\alpha =1..12$ enumerates fermionic components inside
each of $N$ elementary cells. In the Fourier representation the Hamiltonian
reads: 
\begin{equation}
H=\sum_{\mathbf{k}\in K_{+}}\psi _{\mathbf{k}}^{+}\hat{H}_{\mathbf{k}}\psi _{%
\mathbf{k}}  \label{Hferm}
\end{equation}%
where summation is going over the half of the Brilluen zone $K_{+}=\left(
0\leq k_{x}\leq \pi ,~-\pi \leq k_{y}\leq \pi \right) $. Fourier-transformed
Majorana fields, restricted to $K_{+}$, define complex fermions. Hamiltonian 
$H$ is diagonal in the number of this fermions, which results from the
translational invariance of the system. $\hat{H}_{\mathbf{k}}$ is a $%
12\times 12$ gauge-dependent Hermitian matrix. Spectral equation $\det
\left( \hat{H}_{\mathbf{k}}-\epsilon \right) =0$ determines twelve bands
with dispersions $\epsilon _{\alpha ,\mathbf{k}}$. \ The Fermi-sea energy
can then be calculated as $E=\sum f\mathbf{(}\epsilon _{\alpha ,\mathbf{k}}%
\mathbf{)}\epsilon _{\alpha ,\mathbf{k}}$, with $f(\epsilon )$ being the
fermion population numbers (below we imply periodic boundary conditions).
When the ground state of Majorana system in some fixed gauge is found, the 
true ground state of the original spin Hamiltonian should be found
applying the projection operator $P=\prod_{i}\frac{1+D_{i}}{2}$.
However, if we are interested in calculation of 
gauge-invariant quantities (like ground state energy or spin susceptibility), 
there is no need for explicit implementation of this projection,
 and calculation can be done in any particular gauge.

Below we consider vicinity of the point $g=g_{c}=\sqrt{2}$ where the state $%
G_{2}$ becomes critical (see below). Ground state energies of $G_{1..4}$
states per unit cell at $g=g_{c}$ are at $\lambda =0$: $%
E_{1}^{(0)}=-10.758J,~E_{2}^{(0)}=-10.681J,~E_{3}^{(0)}=-10.664J,~E_{4}^{(0)}=-10.610J 
$. That confirms that $G_{1}$ has the lowest global energy at $\lambda =0$.
However, at finite $\lambda $ this energies are simply shifted by $\pm
6\lambda $ and AFM orderdering of $t_{p}$ realized by the $G_{2}$ state
becomes favourable at $\lambda >\lambda _{c}=\left(
E_{2}^{(0)}-E_{1}^{(0)}\right) /12\approx 6.4\times 10^{-3}J$. The $G_{2}$
state breaks both $\mathcal{\hat{T}}$-symmetry and the symmetry $\mathcal{%
\hat{P}}$ of inversion between sublattices of the honeycomb lattice, and it
can be called "chiral AFM" state. However, it is invariant with respect to
the combined inversion $\mathcal{\hat{T}}\mathcal{\hat{P}}$. As temperature
raises above some critical value $T_{c}$, a phase transition leading to a
"chiral-disordered" state obeying both $\mathcal{\hat{T}}$- and $\mathcal{\ 
\hat{P}}$-inversions should occur. We assume below that $T\ll T_{c}$, and
neglect excitations which flip chiralities $t_{p}$.

Eigenstates of (\ref{Hferm}) are found via matrix diagonalization: $\hat{H}_{%
\mathbf{k}}=\hat{S}_{\mathbf{k}}\hat{\tilde{H}}_{\mathbf{k}}\hat{S}_{\mathbf{%
k}}^{+}$. Solving the equation $\det \hat{H}_{\mathbf{k}}=0$ at $g=g_{c}$ we
find that zero-energy excitations are located at two inequivalent points: $%
\mathbf{K}_{1}$ and $\mathbf{K}_{2}$. So, there is a single gapless band $%
\epsilon _{1,\mathbf{k}}$ containing low-energy excitations, whereas all other
11 bands $\epsilon _{2..12,\mathbf{k}}$ have the gap of the order of $J$.
Low-energy excitations are given by $\phi _{1,\mathbf{k}}=\sum_{\alpha
}S_{\alpha }^{+}\psi _{\alpha ,\mathbf{K}_{1}+\mathbf{k}}$ and $\phi _{2,%
\mathbf{k}}=\sum_{\alpha }\bar{S}_{\alpha }^{+}\psi _{\alpha ,\mathbf{K}_{2}+%
\mathbf{k}}$ (hereafter for brevity we write $S_{\alpha }=S_{\alpha ,%
\mathbf{K}_{1}}$ and$~\bar{S}_{\alpha }=S_{1\alpha ,\mathbf{K}_{2}})$.
Perturbation expansion up to the second order in  $\mathbf{k}$
and up to the first order in $\Gamma = g_c - g \ll 1 $ leads to the effective 
Hamiltonian of the low-energy excitations: 
\begin{equation}
H_{eff}=\sum_{|\mathbf{k|\ll }1}\left( \phi _{2,\mathbf{k}}^{+}\phi _{2,%
\mathbf{k}}-\phi _{1,\mathbf{k}}^{+}\phi _{1,\mathbf{k}}\right) \epsilon _{%
\mathbf{k}}\,  \label{Hchi}
\end{equation}%
where $\epsilon _{\mathbf{k}}=\frac{1}{2\sqrt{3}}J\left(
3k_{x}^{2}+k_{y}^{2}\right) -\mu $\, ,  with $\mu =\sqrt{8/3}\,J\,\Gamma $. Density
of states is defined by $\int \frac{d^{2}\mathbf{k}}{\left( 2\pi \right) ^{2}}=$
$\nu \int d\epsilon ~$ and is equal to $\nu =\left( 2\pi J\right) ^{-1}.$
Hamiltonian (\ref{Hchi}) determines low-energy properties of the spin system
(\ref{H}) at small $\Gamma $ and under condition $\lambda >\lambda _{c}$.
The spectrum is gapful at $\Gamma <0$. Positive $\Gamma $ corresponds to the
spin metal state. At $T\ll \mu \ll J$ the heat capacity of spin liquid (per
unit cell) is $C(T)=\frac{\pi }{3}T/J$ ,  demonstrating standard
Fermi-liquid behaviour at low temperatures. However, these gapless excitations
do not carry spin, while spin excitations are
gapped in this model, as we discuss below.

We note that similar analysis could be developed for the $G_3$ state 
in the vicinity of the point $g=\sqrt{3}$, which is known to be critical
for the $G_1$ state as well~\cite{YK}. However, within the model defined by the
 Hamiltonian (\ref{H}), the state $G_3$  always has  energy higher than that of 
 $G_2$ state; the latter, however, has large gap near $g=\sqrt{3}$, thus we 
work near the point $g=\sqrt{2}$ where $G_2$ becomes critical.

Now we turn to the calculation of frequency-dependent spin susceptibility $%
\chi (\omega ,T)$ of the spin metal state. Linear susceptibility tensor is
proportional to the unit matrix due to cubic symmetry of the Hamiltonian (%
\ref{H}) in the spin space. We choose external homogenious field $h(t)$ in
the $z$ direction which adds the term $-h(t)\sum_{\mathbf{r}\alpha }\sigma
_{\alpha }^{z}(\mathbf{r})$ to the spin Hamiltonian and calculate $
\left\langle \sigma ^{z}\right\rangle $. Susceptibility reads $\chi (\omega
)=\sum_{\mathbf{r},\mathbf{r}^{\prime }}\sum_{\alpha \beta }\chi _{\alpha
\beta }(\mathbf{r}^{^{\prime }}-\mathbf{r},\omega )$ with 
\begin{equation}
\chi _{\alpha \beta }(\mathbf{r},~\omega )=i\int_{0}^{\infty }e^{i\omega
t}\left\langle [\sigma _{\alpha }^{z}(\mathbf{r},t),\sigma _{\beta }^{z}(
\mathbf{0},0)]\right\rangle dt  \label{chi}
\end{equation}%
according to Kubo formula ($\mathbf{r}$ enumerates cells and $\alpha ,\beta $
stay for sites within the same cell, average hereafter is taken over the
nonperturbed ground state). We find, following~\cite{Baskaran07}, that
correlation function $G_{\alpha \beta }(\mathbf{r},t)=\left\langle \sigma
_{\alpha }^{z}(\mathbf{r},t)\sigma _{\beta }^{z}(\mathbf{0},0)\right\rangle
_{T}$ is non-zero either for coinciding spins or for spins which are
connected by $z$ or $z^{\prime }$ link. This means, that $G_{\alpha \beta }(
\mathbf{r},t)\sim \delta \left( \mathbf{r}\right) $ (since different
elementary cells are connected by $x^{\prime },~y^{\prime }$ links) and in
what follows we do not write spatial coordinates explicitly. Spin operator
creates two $\mathcal{Z}_{2}$ vortices in the neigbouring plaquettes, which
have an excess energy $\Omega _{\alpha \beta }\sim J$, thus correlation
function $G_{\alpha \beta }(t)$ oscillates with a frequency $\Omega _{\alpha
\beta }$. Therefore dynamic susceptibility $\chi (\omega )=N\sum_{l}\left(
\chi _{\alpha \beta }+\chi _{\beta \beta }+\chi _{\alpha \beta }+\chi
_{\beta \alpha }\right) $, where summation goes over $z$ and \thinspace $
z^{\prime }$ links $l=\left\langle \alpha \beta \right\rangle $ in the unit
cell, contains two resonances at frequencies $\omega =\Omega _{z},\Omega
_{z^{\prime }}$. Our goal is to find lineshapes of these resonances.

The sum in parenthesis in the above expression for $\chi (\omega )$ can be
written as $\chi _{l}=4i\int_{0}^{\infty }e^{i\omega t}\left( G_{l}\left(
t\right) -G_{l}^{\ast }\left( t\right) \right) $ with $G_{l}\left( t\right)
=\left\langle e^{iHt}\psi _{l}e^{-iH_{l}^{\prime }t}\psi
_{l}^{+}\right\rangle $, where $\psi _{l}=\frac{1}{2}\left( c_{\alpha
}+iu_{\alpha \beta }c_{\beta }\right) $  is a complex fermion
defined on a link $l$. In this expression $H_{l}^{\prime }$ stays for
the Hamiltonian  which is different from $H$ by inversion of the sign 
of $u_{l}$: $H_{l}^{\prime }=H+V_{l}$, whereas $V_{l}=4J_{l}\left( \psi _{l}^{+}\psi _{l}-%
\frac{1}{2}\right) $. After standard algebra we find: 
\begin{equation}
G_{l}(t)=\left\langle \psi _{l}(t)T\exp \left( -i\int_{0}^{t}V_{l}(\tau
)d\tau \right) \psi _{l}^{+}\left( 0\right) \right\rangle .  \label{Gl}
\end{equation}%
The problem of calculation of $G_l(t)$ seems to be similar to the Fermi Edge Singularity (FES) problem
with a separable scatterer. The latter was solved exactly 
(in the infrared limit $\mu t \gg 1$)
in \cite{FESzeroT} by summation of the perturbation theory series via the solution
of particular integral equation. However, our problem is, strictly speaking,
different, since initial Hamiltonian $H$ is not diagonal in $\psi
_{l}^{+}\psi _{l}$ and hence correlation function $F_{0l}(t)=\left\langle
T\psi _{l}(t)\psi _{l}\left( 0\right) \right\rangle $ is not equal to zero
identically. However, unlike $G_{0l}(t)=\left\langle T\psi
_{l}\left( t\right) \psi _{l}^{+}\left( 0\right) \right\rangle $ which has
long-time tail $\sim 1/t$, the function $F_{0l}(t\gg \mu ^{-1})$ decays very fast
with $t$ due to exact cancellation between Fermi-surface contributions
coming from different valleys. This means that corresponding pairings in the
series expansion of (\ref{Gl}) do not lead to any singulary at the threshold and
can be neglected. In this case, the solution is similar to the one presented in
Ref.~\cite{FESzeroT} and can be written in terms of the
long-time ($|t\mu |\gg 1$) asymptotics of $G_{0l}\rightarrow -i\nu \left( 
\frac{a_{l}}{t}+\Lambda _{l}\delta \left( t\right) \right) $ (here the term with 
$\delta(t)$ is necessary to preserve the correct weight $\int G_{0l}(t)dt$, see 
\cite{FESzeroT} for details). The $1/t$ tail in the Green function reflects
the presense of a band of gapless excitations $\phi _{1,2}$ which form
Fermi-sea. Reshuffling of the Fermi-sea by the scattering off the local
repulsive potential $V_{l}$ leads to the shift of the energy and to the
change in the power-law exponent in the exact Green function, compared to the
bare one: 
\begin{equation}
G_{l}\left( t\right) =-\frac{i\nu a_{l}}{t}\left( i\xi _{0}t\right)
^{\lambda _{l}}e^{-i\Omega _{l}t},  \label{Glf}
\end{equation}%
where $\lambda _{l}=2\left( \delta _{l}/\pi \right) -\left( \delta _{l}/\pi
\right) ^{2}$ and $\delta _{l}=-\arctan {\frac{4\pi \nu a_{l}J_{l}}{1+4\nu
\Lambda _{l}J_{l}}}$. Frequency $\Omega _{l}\sim J>0$ is the shift in the
ground state energy due to creation of two fluxes. Obviously, $\Omega _{l}$
is different for $z$ and $z^{\prime }$ links; finally, $\xi _{0}\sim \mu $
is the high-energy cutoff. Whereas $a_{l}$ is determined by the vicinity of
Fermi-energy only, the parameter $\Lambda _{l}$ characterizes short-time
behaviour of the Green function, and thus is determined by the whole
spectrum of all 12 fermionic bands: $a_{l}=\frac{1}{N\nu }\sum_{\mathbf{p}
,\gamma }\left( \left\vert S_{\alpha \gamma ,\mathbf{p}}\right\vert
^{2}+\left\vert S_{\beta \gamma ,\mathbf{p}}\right\vert ^{2}\right) \delta
\left( \xi _{\gamma ,\mathbf{p}}\right) $ and $\Lambda _{l}=\frac{2u_{\alpha
\beta }}{N\nu }\sum_{\mathbf{p},\gamma }\xi _{\gamma ,\mathbf{p}}^{-1}\func{
Im}\left[ S_{\alpha \gamma ,\mathbf{p}}S_{\beta \gamma ,\mathbf{p}}^{\ast }
\right] .$ The parameters $a_{l\text{ }}$ and $\Lambda _{l}$ are gauge-independent
constants, which  depend on the type of the link only. Explicite
calculation leads to $a_{l}=\left( \left\vert S_{\alpha }\right\vert
^{2}+\left\vert S_{\beta }\right\vert ^{2}+\left\vert \bar{S}_{\alpha
}\right\vert ^{2}+\left\vert \bar{S}_{\beta }\right\vert ^{2}\right) $ and $
\Lambda _{l}=c_{l}\log \left( \frac{J}{\mu }\right) +\eta _{l}$, where $
c_{l}=2u_{\alpha \beta }\func{Im}\left[ \bar{S}_{\alpha }\bar{S}_{\beta
}^{\ast }-S_{\alpha }S_{\beta }^{\ast }\right] $ and $\eta _{l}$ is some
number of the order of unity which can be found only by numerical
integration over $K_{+}$ (it is determined by the whole band). Evaluation
leads to the following result:
\begin{equation}
\tan \delta _{z}=\frac{-1}{1.04+0.18\ln \frac{J}{\mu }},~\tan \delta
_{z^{\prime }}=\frac{1}{0.40+0.26\ln \frac{J}{\mu }}.  \label{delta-f}
\end{equation}%
Eq.(\ref{delta-f}) determines phase shifts modulo $\pi $ only. 
This ambiguity is fixed by the continuity condition: $\delta _{l}(\mu
\rightarrow 0)=0$. Since $\lambda _{z}<0$ and $\lambda _{z^{\prime }}>0$ for
any $J/\mu $, only $\chi _{z^{\prime }}$ diverges at the corresponding
threshold (while $\chi _{z}$ still has a cusp); therefore  below we concentrate on the
contribution of the $z^{\prime }$ links only, $\lambda _{z^{\prime }}\equiv
\lambda $.
\begin{figure}[tbp]
\includegraphics[width=8cm,height=5cm]{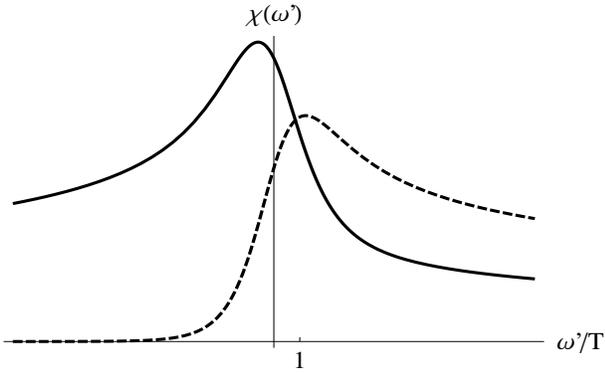}
\caption{Real (full line)  and imaginary (dashed line) parts of 
$\protect\chi (\protect\omega ^{\prime
},T)$ at $\protect\mu /J=0.05$, that corresponds to the exponent $\protect
\lambda =0.35$.}
\label{chi-fig}
\end{figure}
Using the results (\ref{Glf},\ref{delta-f}), we calculate spin
susceptibility close to the $\Omega _{z^{\prime }}$ resonanse, with $\omega
^{\prime }=\omega -\Omega _{z^{\prime }}$. We find $\chi _{T=0}\left( \omega
^{\prime }\right) =\chi _{0}\Gamma \left( \lambda \right) \left( -\xi
_{0}/\omega ^{\prime }\right) ^{\lambda },$ where $\chi _{0}=\frac{4N}{3\pi }
J^{-1}.$ Note, that $\Im \chi _{l}=0$ below the threshold ($\omega ^{\prime
}<0)$ as it should be at $T=0$. These results can be easily generalized to
the finite temperature $T\ll \mu $. As was shown in \cite{FESfinT},
finite-temperature correlation function in the FES\ problem can be obtained
from the zero-temperature one by substitution $t\rightarrow \frac{\sinh \pi
Tt}{\pi T}$. For the susceptibility, that gives: 
\begin{equation}
\chi \left( \omega ^{\prime }\right) =\chi _{0}\left( \frac{i\xi _{0}}{2\pi T%
}\right) ^{\lambda }\frac{\Gamma \left( \lambda \right) 
\Gamma \left( \frac{1-\lambda}{2}  -\frac{i\omega }{2\pi T}\right) }
{\Gamma \left( \frac{1+\lambda}{2} -\frac{i\omega }{2\pi T}\right) }.  
\label{chi_f}
\end{equation}%
This function is plotted in Fig.~\ref{chi-fig}. 
The major effect of nonzero temperature is the appearence of absorption
below threshold: $\chi \left( -\omega ^{\prime }\gg T\right) \approx \chi
_{T=0}\left( \omega ^{\prime }\right) \left( 1+i\sin \pi \lambda e^{-|\omega
^{\prime }|/T}\right) $; in addition, the resonant peak appears to be
smeared out: $\chi \left( |\omega ^{\prime }|\ll T\right) \approx \chi
_{0}e^{i\pi \lambda }\left( \frac{\xi _{0}}{2\pi T}\right) ^{\lambda }$.

In conclusions, we have shown that certain (numerically, very weak)
modification of the Yao-Kivelson version of Kitaev spin lattice leads to the
ground-state of the Fermi-liquid type, with a Fermi energy $\mu \propto 
\sqrt{2}-J^{\prime }/J$. We have studied the model in the continuum limit of
small Fermi-circle $\mu \ll J$ and at low temperatures $T\ll \mu $. Gapless
excitations of the Fermi-sea do not carry spin themselves, but they
determine the shape (\ref{chi_f}) of the resonance peak in the dynamic spin
susceptibility.

We are grateful to A. Yu. Kitaev for numerous important discussions and advises.
This research was supported by the RFBR grant \# 10-02-00554 and by the RAS
Program "Quantum physics of condensed matter".

\end{document}